\documentclass[12pt,preprint]{aastex}

\shorttitle{SiO in the H211 protostellar jet}
\shortauthors{Hirano et al.}

\received{2005 May 1}
\begin{document}

\title{SiO $J$=5--4 in the HH\,211 protostellar jet imaged with the SMA}

\author{NAOMI HIRANO\altaffilmark{1,2}, SHENG-YUAN LIU\altaffilmark{1}, HSIEN SHANG\altaffilmark{1},
PAUL T.P. HO\altaffilmark{1,3}, HUI-CHUN HUANG\altaffilmark{4},
YI-JEHNG KUAN\altaffilmark{4,1}, MARK J. McCAUGHREAN\altaffilmark{5},
\& QIZHOU ZHANG\altaffilmark{3}}

\altaffiltext{1}{Academia Sinica,
Institute of Astronomy \& Astrophysics, P.O. Box 23--141, Taipei, 106, 
   Taiwan, R.O.C.}
\altaffiltext{2}{e-mail: hirano@asiaa.sinica.edu.tw}

\altaffiltext{3}{Harvard-Smithsonian Center for Astrophysics, 60 Garden
   Street, Cambridge, MA 02138, USA}

\altaffiltext{4}{Department of Earth Science, National Taiwan Normal 
   University, 88 Sec.\ 4, Ting-Chou Rd., Taipei, 116, Taiwan, R.O.C.}

\altaffiltext{5}{School of Physics, University of Exeter,
   Stocker Road, Exeter EX4 4QL, Devon, UK}

\begin{abstract}
We have mapped the SiO $J$=5--4 line at 217\,GHz from the HH\.211 molecular 
outflow with the Submillimeter Array (SMA). The high resolution map 
(1.6$''{\times}$0.9$''$) shows that the SiO $J$=5--4 emission comes from the 
central narrow jet along the outflow axis with a width of $\sim$0.8$''$ 
($\sim$\,250\,AU) FWHM\@. The SiO jet consists of a chain of knots separated 
by 3--4$''$ ($\sim$\,1000\,AU) and most of the SiO knots have counterparts in 
shocked H$_2$ emission seen in a new, deep VLT near-infrared image of the 
outflow. A new, innermost pair of knots are discovered at just $\pm$2$''$ from
the central star. The line ratio between the SiO $J$=5--4 data and upper limits 
from the SiO $J$=1--0 data of \citet{Cha01} suggests that these knots have a 
temperature in excess of 300--500\,K and a density of 
(0.5--1)$\times$10$^7$\,cm$^{-3}$.
The radial velocity measured for these knots is $\sim$\,30\,km\,s$^{-1}$, 
comparable to the maximum velocity seen in the entire jet. The high 
temperature, high density, and velocity structure observed in this pair of 
SiO knots suggest that they are closely related to the primary jet launched 
close to the protostar.
\end{abstract}


\section{INTRODUCTION}
Several
highly-collimated molecular outflows driven by deeply-embedded young stellar 
objects are known to have an extremely-high velocity (EHV) flow component, with terminal 
velocities of 50--150\,km\,s$^{-1}$\citep[e.g.,][]{Bac96}. These EHV 
components are closely confined to the axes of the lobes and have large 
momenta, comparable to those of the slowly-moving (20--30\,km\,s$^{-1}$) 
``classical'' outflows. As a result, it is felt that the EHV flow is closely 
connected with the ``primary jet'' responsible for driving the broader 
molecular outflow.

The HH\,211 outflow (D\,$\sim$\,315\,pc)
was discovered via near-infrared H$_2$ imaging \citep{McC94} and is an 
archetypal outflow with a highly-collimated EHV jet. The outflow is 
driven by a low-luminosity (3.6\,$L_{\odot}$) Class~0 protostar 
($T_{\rm bol}$\,$\sim$\,33\,K) and is thought to be extremely young 
($\tau_{\rm dyn}$\,$\sim$\,750\,yr). The CO $J$=2--1 images of
\citet[][hereafter GG99]{Gue99} with a spatial resolution of 1.5$''$ show remarkable features: 
the low-velocity CO delineates a pair of cavities whose tips are associated 
with the near-infrared H$_2$ emission, while the high-velocity CO traces a 
narrow jet whose velocity increases linearly with distance from the star.
In spite of a small opening angle of 22$^{\circ}$, the CO $J$=2--1 emission 
from the cavity appears on both sides of the central star at velocities
close to the systemic velocity of $V_{\rm LSR}$\,=\,9.2\,km\,s$^{-1}$ 
\defcitealias{Gue99}{GG99}\citepalias{Gue99}, implying that the outflow axis lies within 10$^\circ$ of the
plane of the sky. The jet component (but not the cavity) is also traced by 
thermal SiO emission, in $J$=1--0 \citep[][hereafter CR01]{Cha01}, $J$=5--4 \citep{Gib04},
$J$=8--7 and $J$=11--10 \citep{Nis02} transitions. Since these lines have 
critical densities larger than ${\sim}10^6$\,cm$^{-3}$ and the energy level 
of $J$=11 is higher than 100\,K, the detection of these SiO lines means that 
the jet is much denser and warmer than the lower-velocity cavity component.

\section{OBSERVATIONS}
The observations of the SiO $J$=5--4 transition at 217.105\,GHz were carried 
out on January 30, 2004 with the six antennas of the Submillimeter Array 
(SMA)\footnote{The Submillimeter Array (SMA) is a joint project between the 
Smithsonian Astrophysical Observatory and the Academia Sinica Institute of 
Astronomy and Astrophysics, and is funded by the Smithsonian Institution and 
the Academia Sinica.} on Mauna Kea, Hawaii \citep{Ho04}. We used an extended 
array configuration that provides baselines ranging from 17\,m to 179\,m.
The primary-beam size (HPBW) of the 6\,m diameter antennas at 217\,GHz was 
measured to be $\sim$54$''$. Thus, in order to cover the whole outflow 
($\sim$\,90$''$ along the major axis), we observed three fields separated 
by 25$''$ along the major axis of the outflow (Fig.\ 1). The spectral 
correlator was set to provide a uniform 
frequency resolution of 406.25\,kHz across a 2\,GHz wide band.
The visibility data were calibrated using the MIR package, with
3C84 used as a phase and amplitude calibrator, and Ganymede and Titan as 
flux calibrators: uncertainty in the flux scale is estimated at 
$\sim$\,20\%. The bandpass was calibrated by observations of Jupiter or 
Saturn. The calibrated visibility data were imaged using MIRIAD, followed by
a non-linear joint deconvolution using the CLEAN-based algorithm, MOSSDI,
with robust weighting of 0.5 providing a synthesized beam of 
1.6$''{\times}$0.9$''$ at a position angle of $-$41$^{\circ}$. A continuum 
map was obtained by averaging the line-free channels of both sidebands 
separated by 10\,GHz, this time with natural weighting providing a 
synthesized beam of 1.7$''{\times}$0.9$''$ at a position angle of 
$-$44$^{\circ}$. 
Figure 1 shows a compact 220\,GHz continuum source
at $\alpha$ = 03$^h$43$^m$56.79$^s$, $\delta$ = 32$^{\circ}$00$'$50.0$''$ 
(J2000.0) with a peak flux of $\sim$\,73\,mJy beam$^{-1}$ (12$\sigma$) and a total integrated flux of $\sim$\,200\,mJy. This latter flux is approximately 
70\% of the 230\,GHz flux measured by \citetalias{Gue99} using similar baselines 
(24--180\,m) to ours, and it is likely that the relatively poor sensitivity 
of our measurements ($1\sigma$\,$\sim$\,6.2\,mJy\,beam$^{-1}$) failed to 
recover all of the low-level emission seen in the higher sensitivity 
($1\sigma$\,$\sim$\,3\,mJy\,beam$^{-1}$) data of \citetalias{Gue99}. 

A new, deep near-infrared image of
HH\,211 was obtained using the ISAAC
instrument of the ESO Very Large Telescope at Paranal, Chile on January~4,
2002 under clear conditions.
Images were taken through a 1\% wide filter centered on the v=1--0 S(1)
line of H$_2$ at 2.122\,$\mu$m. A 2$\times$2 position mosaic 
was used to cover a 5$\times$5 arcmin region centered near HH\,211-mm at an image scale of 0.148 arcsec/pixel. In the 
central, fully-overlapping region, the total integration time was 
8\,min\,pixel$^{-1}$ after mosaicing.
The data reduction was standard, including the subtraction of a blank sky
image made from a median-filtered stack of the source images and flat fielding
using a sequence of twilight-illuminated images. The data were corrected for
the small geometric distortion in ISAAC using the coefficients determined from images of M\,16 by \citet{McC02},
before being aligned and mosaiced. Despite a mean airmass of 1.9, the 
resulting resolution is 0.60 arcsec FWHM and accurate astrometry (0.10 arcsec 
rms) was derived from 17 stars in common with the 2MASS Point Source Catalog. 
An approximate flux calibration yields a 3$\sigma$ surface brightness
limit of $1.4\times 10^{-20}$\,W\,m$^{-2}$\,pixel$^{-1}$. 
The central portion of this new
image shown in greyscale in Figure~1 well delineates the key structures in the
outflow, including the knots along the central axis and the excavated
cavity (part of which is continuum emission as shown in Figure 1 of \citealp{Eis03}) outside them. 

\section{RESULTS AND DISCUSSION}
\subsection{The highly-collimated SiO jet}
SiO $J$=5--4 emission was detected from HH\,211 in two velocity ranges from 
$-$24km\,s$^{-1}$ to $-$4\,km\,s$^{-1}$ (blueshifted) and from 
+4\,km\,s$^{-1}$ to +32\,km\,s$^{-1}$ (redshifted) with respect to the 
systemic velocity of $V_{\rm LSR}$\,=\,9.2\,km\,s$^{-1}$. Integrated intensity 
maps of the blueshifted and redshifted emission are shown in Figure~1 and
velocity channel maps at 4\,km\,s$^{-1}$ intervals are presented in Figure~2.
The SiO $J$=5--4 emission is seen to be concentrated exclusively in the 
jet-like narrow region along the outflow axis and there is no counterpart 
to the low-velocity cavity component seen in CO $J$=2--1 \citepalias{Gue99}. We 
have smoothed the SMA map to a resolution of 22$''$ and compared it with the 
SiO $J$=5--4 spectrum observed with the JCMT \citep{Gib04}, finding that 
80--100\% of the single-dish flux is recovered by the SMA, despite the 
missing short spacing information. This confirms that almost all of the 
SiO $J$=5--4 emission arises from a narrow jet.
After deconvolution of the SMA beam, the width of the SiO $J$=5--4 jet is 
$\sim$0.8$''$ ($\sim$\,250\,AU) FWHM\@. It comprises a chain of knots 
each separated by $\sim$\,3--4$''$ ($\sim$\,1000\,AU), with five
discrete knots (R1--R5 in Fig.~2) on the redshifted side and six (B1--B5)
on the blueshifted side. Knots B2--B5, R3, and R4 all appear to have 
near-infrared H$_2$ counterparts, while the lack of H$_2$ emission coincident 
with B1, R1, and R2 is probably due to the very high extinction associated 
with the dense protostellar envelope as traced in H$^{13}$CO$^+$ \citepalias{Gue99} 
and NH$_3$ \citep{Wis01}. In addition, the innermost knots, B1 and R1, have 
no counterparts in SiO $J$=1--0 \defcitealias{Cha01}{CR01}\citepalias{Cha01} or CO $J$=2--1 \citepalias{Gue99}.

The blueshifted side of the SiO $J$=5--4 jet ends near an H$_2$ knot roughly 
23$''$ to the east of the protostar and the redshifted side 
$\sim$\,18$''$ to the west of the center. 
The SiO $J$=5--4
does not not extend as far as the CO $J$=2--1 jet observed by \citetalias{Gue99}, 
terminating at features BI and RI (marked on Figure 1), respectively: no SiO $J$=5--4 is seen 
beyond these positions, even though our three-pointing mosaic covered almost 
the entire outflow. The lowest transition SiO $J$=1--0 also terminates at BI 
and RI \citepalias{Cha01}, and the implication is that beyond these points, the jet 
is no longer dense enough to excite SiO emission. An alternative hypothesis is 
that beyond BI and RI, the SiO has been destroyed via chemical processing in 
shocks, as described by \citetalias{Cha01}.

An SiO $J$=5--4 position-velocity diagram along the jet axis (Figure~3) shows 
a different velocity structure to that seen in CO $J$=2--1 \citepalias{Gue99}. In
particular, the SiO $J$=5--4 exhibits a large velocity dispersion at the 
innermost knots B1 and R1, suggesting that these knots are kinematically 
distinct from those in the outer part. At $\sim$\,2$''$ ($\sim$\,630\,AU) 
from the protostar, the velocity centroid and the terminal velocity reach 
$\pm$\,18\,km\,s$^{-1}$ and ${\sim}{\pm}$\,30\,km\,s$^{-1}$, respectively .
The outer part of the SiO $J$=5--4 shows a Hubble-like velocity structure,
previously seen for the CO $J$=2--1. However, the SiO jet is moving 
$\sim$\,5\,km\,s$^{-1}$ faster than the CO jet, reaching a maximum radial
velocity of $\sim$\,35\,km\,s$^{-1}$. Assuming that the axis of the flow is 
inclined by $\sim$\,10$^{\circ}$ out of the plane of the sky, the deprojected 
outflow velocity in the jet would correspond to $\sim$\,200\,km\,s$^{-1}$, 
typical of the velocity of the primary jet driven by low-mass stars.

In Figure~4, we compare line profiles of the SiO $J$=5--4 at knots B1, R1, 
B4, and R3, with those observed in $J$=1--0 with the VLA by \citetalias{Cha01}.
In order to obtain line intensity ratios (Table~1), we convolved the data 
cubes of the two transitions to an equal angular resolution of 1.6$''$.
Then, to derive physical parameters in the jet as traced by SiO emission, we 
carried out large velocity gradient (LVG) statistical equilibrium calculations 
following the method of \citet{Nis02}. \citetalias{Cha01} reported peak brightness 
temperatures of the SiO $J$=1--0 of 54\,K and 47\,K in the redshifted and
blueshifted gas, respectively, averaged over a 0.58$''{\times}$0.43$''$ beam.
These SiO $J$=1--0 emission peaks correspond to the positions of R3 and B4
in our data, where the 5--4/1--0 ratio is seen to be $\sim$\,2. The LVG 
calculations suggest that a density of (0.5--1)$\times$10$^{7}$\,cm$^{-3}$, 
a temperature of 60--120\,K, and an SiO abundance $X$(SiO) of 
(0.5--1)$\times$10$^{-6}$ are required to satisfy both the lower limit of 
the SiO $J$=1--0 brightness temperature and the 5--4/1--0 ratio. As no
$J$=1--0 emission was detected at the innermost knots, B1 and R1, the
5--4/1--0 ratio must exceed 10: in order to reproduce such ratios, 
the same density range of (0.5--1)$\times$10$^{7}$\,cm$^{-3}$ is again
required, but at a much higher gas kinetic temperature, in excess of 
300--500\,K\@. These density and temperature ranges for B1 and R1 agree with 
those derived from the higher-$J$ transition SiO lines observed with a 
single-dish telescope \citep{Nis02} and the high-$J$ CO lines observed 
with the ISO Long Wavelength Spectrometer \citep{Gia01}.

\subsection{The structure of the HH211 outflow}
The rim-brightened shapes seen in the near-infrared and CO $J$=2--1 images 
suggest that the outflow lobes are the cavities filled with lower-density gas.
Since the mean density of the molecular cloud core in which HH\,211 lies is 
$\sim$\,4$\times$10$^4$\,cm$^{-3}$ \citep{Bac87}, the density inside the 
cavities should be on the order of 10$^{4}$\,cm$^{-3}$ or less. On the other 
hand, the SiO jet along the lobe axis has a much higher density of 
(0.5--1)$\times$10$^{7}$\,cm$^{-3}$. The strong H$_2$ emission observed in 
HH\,211 suggests that the shocks in this outflow are nondissociative C-shocks 
\citep[e.g.,][]{Hol97} and therefore, it is unlikely that this emission is
coming from cavity gas which has been compressed by a factor of $>$500 from
the ambient $\sim$\,10$^4$\,cm$^{-3}$. Rather, it is much more likely that
the primary jet itself is has an intrinsically high density or that dense 
gas in the protostellar disk or envelope is being entrained by the primary 
jet. While the FWHM of the dense SiO $J$=5--4 jet is $\sim$\,0.8$''$, the
CO $J$=2--1 jet was seen to be somewhat broader at $\sim$\,1.5$''$ near the
protostar and increasing up to $\sim$\,3$''$ further out, and with a density
of $\sim$\,10$^5$\,cm$^{-3}$, roughly 50 times lower than seen in the SiO
\citepalias{Gue99}. These results suggest that the jet has an axial structure,
with the higher density ($>$10$^6$\,cm$^{-3}$) gas close to the axis surrounded 
by a sheath of less dense gas 10$^5$\,cm$^{-3}$.

The Hubble law velocity structure seen in Figure 3 at $> \pm 5''$ from the source is predicted from shells driven by wide-angle winds \citep[e.g.,][]{Shu91, Lee01}.
On the other hand, turbulent entrainment models require an acceleration region near the star followed by a deceleration region \citep[e.g.,][]{Rag93}, and bow shock entrainment models show a broad range of velocities near the bow tips \citep[e.g.,][]{Lee01}; these do not explain the Hubble low shown in Figure 3.
The highly-collimated morphology of the SiO $J$=5--4 jet can be explained as an on-axis density enhancement within the $X$-wind type of wide-opening angle wind \citep{Li96, Sha02}.
Since the wide-angle wind model has large radial velocity component, this will explain the 
large velocity dispersion observed near the base (at B1 and R1) combined with the jet-like density enhancement, which cannot be explained by any other entrainment models. 

The density and velocity structures observed with the SMA suggest that the 
highly-collimated jet traced by the SiO $J$=5--4 is closely related to the 
primary jet. In particular, the innermost pair of knots, B1 and R1, have a 
high density of $>$10$^6$\,cm$^{-3}$, a high temperature of $>$300--500\,K, 
and a large velocity dispersion  of 30\,km\,s$^{-1}$, and quite plausibly
represent the primary jet itself, immediately after being launched from the 
protostar/disk system.

\acknowledgments

We wish to thank all the SMA staff in Hawaii, Cambridge, and Taipei for their 
enthusiastic help during these observations. We acknowledge T. Oka for 
consulting on the LVG calculation, K. Dobashi for helping us to overlay the 
SiO and H$_2$ images, and C. Chandler for providing us with the published
VLA SiO $J$=1--0 data.

\clearpage

\begin{deluxetable}{rcrrrrrr}
\tablecolumns{6}
\tablewidth{0pc}
\tablecaption{SiO line intensity values at 1.6$''$ resolution and line ratios 
measured at different offsets from the center along position angle of 
116$^{\circ}$}
\tablehead{
\colhead{}    & \colhead{}    
& \multicolumn{2}{c}{Integrated intensity (K\,km\,s$^{-1}$)} &   \colhead{}   &
\colhead{} \\
\cline{3-4} \\
\colhead{Offset} & \colhead{knot}   & \colhead{$J$=5--4}    & \colhead{$J$=1--0}    & \colhead{}   & \colhead{I(5--4)/I(1--0)}}
\startdata
$-$12$''$ & R3  & 342.9$\pm$14.4 &
148.6$\pm$8.3 & &2.31$\pm$0.16 \\
$-$1$''$ & R1 & 232.4$\pm$14.4 &
23.9$\pm$8.3 & & 9.72$\pm$3.43 \\
$+$1$''$ & B1 & 189.9$\pm$13.4 & $<$22.5\phn &
& $>$8.44\phn \\
$+$13$''$ & B4 & 163.8$\pm$13.4 & 101.9$\pm$7.5 &
& 1.67$\pm$0.18 \\
\enddata
\end{deluxetable}

\clearpage

\begin{figure}
\epsscale{0.75}
\plotone{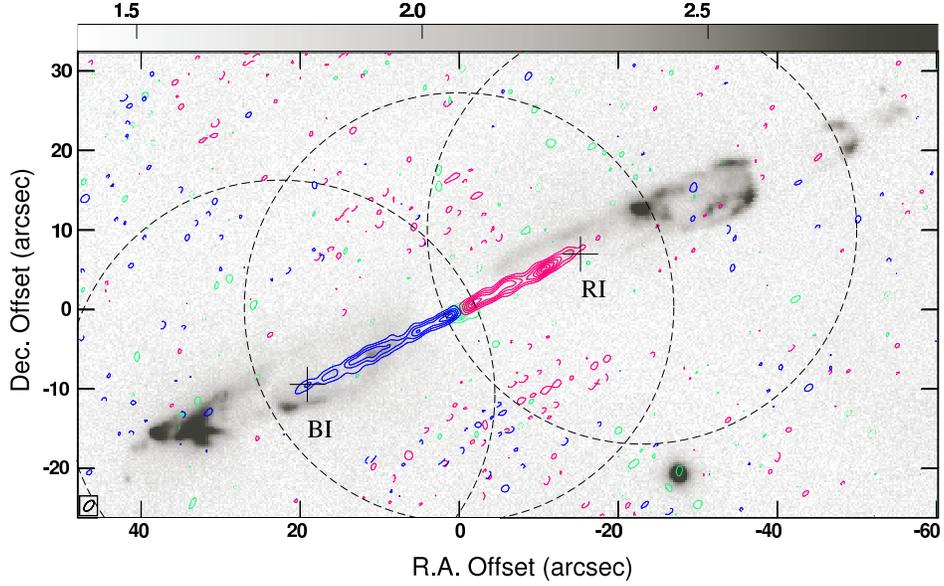}
\caption{Integrated blue (blue contours) and redshifted (red contours) SiO 
$J$=5--4 and 220\,GHz continuum (green contours) maps superposed on the 
H$_2$ $v$=1--0 S(1) emission (greyscale). Offsets (in arcsec) are with 
respect to the millimeter continuum position, $\alpha = 03^h43^m56.8^s$, 
$\delta = 32^{\circ}00'50.4''$ (J2000.0) given by \citetalias{Gue99}. The velocity 
range of the blueshifted component is from $-$12.8 to +7.2\,km\,s$^{-1}$, and 
of the redshifted component from +15.2 to +43.2\,km\,s$^{-1}$. The contours 
are drawn at intervals of 2$\sigma$ with the lowest contours at 3$\sigma$.
The 1$\sigma$ values are 1.5\,Jy\,beam$^{-1}$\,km\,s$^{-1}$ for the 
blueshifted component, 2.0\,Jy\,beam$^{-1}$\,km\,s$^{-1}$ for the redshifted 
component, and 6.2\,mJy\,beam$^{-1}$ for the 220\,GHz continuum. The dashed 
circles indicate the half-power primary beams of the three observed fields.
The crossses denote the positions of BI and RI identified in the CO $J$=2--1 
jet by \citetalias{Gue99}.
\label{fig1}}
\end{figure}

\clearpage 

\begin{figure}
\plotone{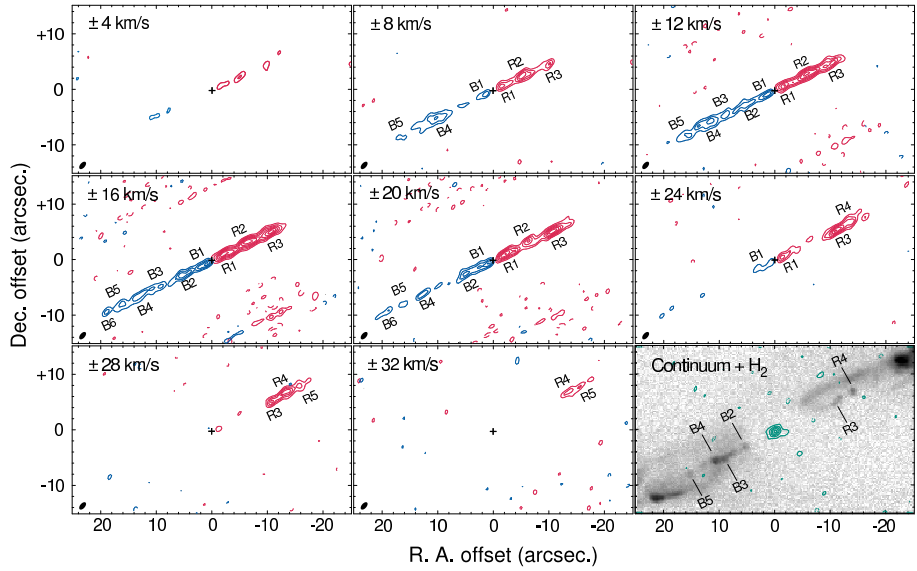}
\caption{Velocity channel maps of the SiO $J$=5--4 emission with a velocity 
interval of 4\,km\,s$^{-1}$. The contour interval is 0.23\,Jy\,beam$^{-1}$ 
(2$\sigma$) with the lowest contour at 0.325\,Jy\,beam$^{-1}$ (3$\sigma$).
The last panel shows the 220\,GHz continuum emission superposed on the 
H$_2$ v=1--0 S(1) image. The contours spaced by 12.4\,mJy\,beam$^{-1}$ 
(2$\sigma$) with the lowest contour at 18.6\,mJy\,beam$^{-1}$ (3$\sigma$).
\label{fig2}}
\end{figure}

\clearpage 

\begin{figure}
\plotone{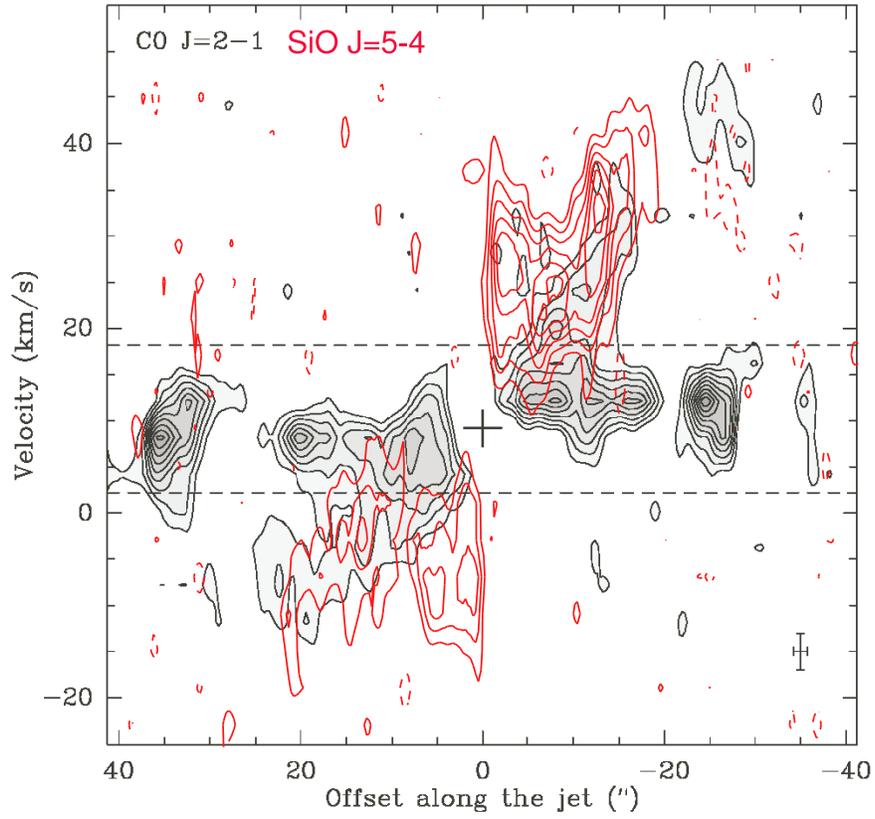}
\caption{Position-velocity map of the SiO $J$=5--4 emission along the outflow 
axis (red contours) superposed on that of the CO $J$=2--1 from \citetalias{Gue99}.
The contour step of the SiO $J$=5--4 data is 15\% of the maximum 
(2.18\,Jy\,beam$^{-1}$).
\label{fig3}}
\end{figure}

\clearpage 

\begin{figure}
\plotone{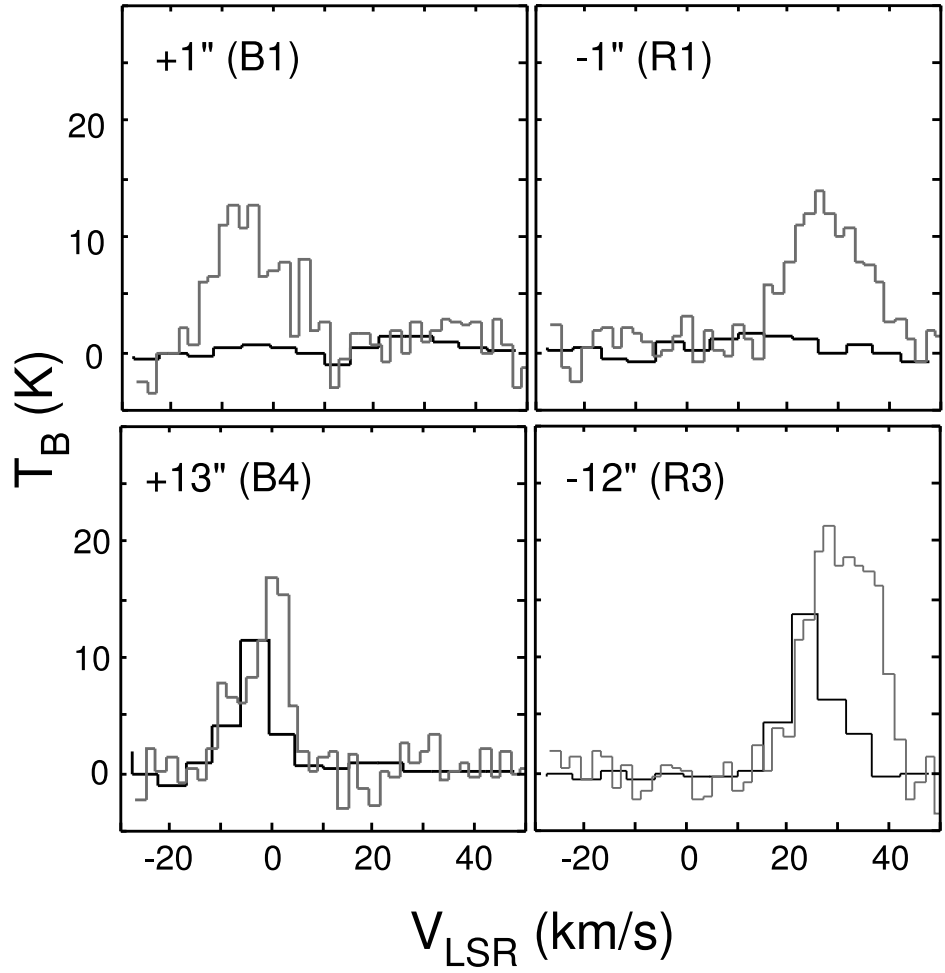}
\caption{Spectra of the SiO $J$=5--4 (grey lines) and $J$=1--0 (black lines) 
measured at different positions along the outflow axis, with both datasets
convolved to 1.6$''$ resolution.
\label{fig4}}
\end{figure}

\end{document}